\documentclass[amssymb,prl,twocolumn,notitlepage,longbibliography,superscriptaddress]{revtex4-1}

\renewcommand{\vec}[1]{\mathbf{#1}}

\newcommand{\braket}[2]{\langle #1 | #2 \rangle}

\newcommand{\ket}[1]{| #1 \rangle}
\newcommand{\bra}[1]{\langle #1|}

\usepackage[%
colorlinks=true,
urlcolor=blue,
linkcolor=blue,
citecolor=blue
]{hyperref}

\usepackage{color}

\usepackage{amsmath}                          
\usepackage{amsfonts}                         
\usepackage{amssymb}                          
\usepackage{amsthm}                           
\usepackage{mathdots}                         

\usepackage{graphicx}                         
\usepackage{subfigure}                        
\usepackage{overpic}

\begin{document}


\title{Protocol for implementing quantum nonparametric learning with trapped ions}

\author{Dan-Bo Zhang}
\affiliation{Guangdong Provincial Key Laboratory of Quantum Engineering and Quantum Materials, GPETR Center for Quantum Precision Measurement, SPTE and SPTE and Frontier Research Institute for Physics\\
South China Normal University, Guangzhou 510006, China}

\author{Shi-Liang Zhu} \email{slzhu@nju.edu.cn}
\affiliation{Guangdong Provincial Key Laboratory of Quantum Engineering and Quantum Materials, GPETR Center for Quantum Precision Measurement, SPTE and SPTE and Frontier Research Institute for Physics\\
	South China Normal University, Guangzhou 510006, China}
\affiliation{National Laboratory of Solid State Microstructures, School of Physics, Nanjing University, Nanjing 210093, China}

\author{Z. D. Wang}
\email{zwang@hku.hk}
\affiliation{Department of Physics and Center of Theoretical and Computational Physics, The University of Hong Kong, Pokfulam Road, Hong Kong, China}
\affiliation{Guangdong Provincial Key Laboratory of Quantum Engineering and Quantum Materials, GPETR Center for Quantum Precision Measurement, SPTE and SPTE and Frontier Research Institute for Physics\\
	South China Normal University, Guangzhou 510006, China}

\begin{abstract}
Nonparametric learning is able to make reliable predictions by extracting information from similarities between a new set of input data and all samples. Here we point out a quantum paradigm of nonparametric learning which offers an exponential speedup over the sample size. By encoding data into quantum feature space, similarity between the data is defined as an inner product of quantum states. A quantum training state is introduced to superpose all data of samples, encoding relevant information for learning in its bipartite entanglement spectrum. We demonstrate that a trained state for prediction can be obtained by entanglement spectrum transformation, using quantum matrix toolbox. We further work out a feasible protocol to implement the quantum nonparametric learning with trapped ions, and  demonstrate the power of quantum superposition for machine learning.
\end{abstract}

\maketitle

\emph{Introduction.--}
Machine learning extracts useful information from data for prediction. The extraction can be categorized into parametric and nonparametric learning~\cite{christopher_m_bishop_pattern_2006,trevor_hastie_elements_2009}.
Parametric learning distills knowledge of data into parameters of a function, e.g., neural networks. However, the form of function may set a model bias or a limitation.  Without a predetermined form of a function, nonparametric learning can make predictions by extracting information of similarities between new data and all samples, with the appropriate sample weighting related to correlation of samples. This can utilize a self-defined kernel that may better capture the similarity between data, while on the other hand, it requires a large number of samples and the runtime is polynomial with the sample size, which is time-consuming for big data.

In quantum setting, machine learning can be enhanced with quantum information  processing~\cite{biamonte_17,das_19,harrow_09,wiebe_12,lloyd_14,rebentrost_14,dunjko_16,lloyd_16,lloyd_18,havlíček_19,schuld_19}. While quantum algorithms of nonparametric learning were studied for Gaussian processes~\cite{das_18,zhao_19,zhao_19-c8Wsa,zhao_19-frAQB}, we focus on  more general cases of nonparametric learning and its enhancement
by exploiting quantum advantages. First, encoding classical data $\vec{x}$ into quantum state $\ket{\psi_{\vec{x}}}$ can take advantages of quantum-enhanced feature spaces for highly nonlinear feature map~\cite{havlíček_19,mitarai_18,schuld_19}, which is desirable for complicated machine learning tasks.
Second, all data of samples can be superposed, and querying of similarities can be achieved in a quantum parallel way. Moreover, correlations of data can be extracted and transformed more efficiently with quantum matrix toolbox~\cite{harrow_09,lloyd_14,Gilyén_2018}, including density matrix exponentiation and matrix inversion.  
	

In this Letter, we illustrate a quantum paradigm for nonparametric learning by elaborating on a regression task and its physical implementation. With a superposition of all samples into a quantum training state $\ket{\psi_\vec{A}}$ defined later ~\cite{schuld_16}, we show that relevant important information for learning is represented by the bipartite entanglement spectrum of $\ket{\psi_\vec{A}}$~\cite{zhang_19}, and different kinds of regression can be proposed by choosing different types of entanglement spectrum transformation. The transformation involves quantum algorithm for matrix inversion using auxiliary qumodes (continuous variables)~\cite{lau_17,zhang_19}.
We further propose a feasible scheme to implement this quantum nonparametric learning with trapped ions
~\cite{leibfried_03,häffner_08,monroe_13},  
and demonstrate the power of quantum superposition for machine learning. Our work provides a new insight for machine learning by exploiting entanglement structure of quantum superposed training data.

\emph{Nonparametric regression.--}
Let us first introduce nonparametric learning.
Given a training dataset of $M$ points  $\{\mathbf{x}^{(m)},y^{(m)}\}$ (with $m=1,2,\cdots,M$), where $\mathbf{x}^{(m)} \in R^N$ is a vector of $N$ features and $y^{(m)}\in R$ is the target value, the goal is to learn an input-output function, which can be used to predict $\tilde{y}$ for new data $\tilde{\mathbf{x}}$.  A parametric regression is to find a function $f(\mathbf{x})$, e.g., a linear model, $f(\mathbf{x})=\mathbf{w}^T\mathbf{x}$, parametrized by a matrix $\mathbf{w}$.
A nonparametric learning, instead, directly establishes a prediction based on a weighted average over the similarity between new data $\tilde{\mathbf{x}}$ and each training data, namely,
 \begin{equation}\label{eq:prediciton}
\tilde{y}=\sum_{m=1}^{M}\alpha_m\kappa(\mathbf{x}^{(m)},\tilde{\mathbf{x}}),
\end{equation}
where $\kappa(\mathbf{x}^{(m)},\tilde{\mathbf{x}})$ defines the similarity between data and can be chosen beforehand. The weighting $\boldsymbol{\alpha}=(\alpha_1,...,\alpha_M)^T$, for instance, can be determined by minimizing the least-square loss function
\begin{equation}\label{eq:least-square}
L(\boldsymbol{\alpha})=\sum_{m=1}^{M}(\tilde{y}^{(m)}-y^{(m)})^2+\chi\sum_{m=1}^{M}\alpha_m^2.
\end{equation}
Here the $\chi$-term is a $L_2$ regularization term that makes a constraint on the weighting of each sample, which is necessary for avoiding over fitting. The combination of Eq.~\eqref{eq:prediciton} and Eq.~\eqref{eq:least-square} is a kernel ridge regression.  The solution turns to be $\boldsymbol{\alpha}=(K+\chi I)^{-1}\vec{y}$, where $K$ is the covariance matrix with elements $K_{m_1,m_2}=\kappa(\mathbf{x}^{(m_1)},\mathbf{x}^{(m_2)})$, and $\vec{y}=(y^{(1)},...,y^{(M)})^T$. The prediction can be written as $\tilde{y}=\vec{y}^T(K+\chi I)^{-1}\boldsymbol{\kappa}$, where $\kappa_m=\kappa(\mathbf{x}^{(m)},\tilde{\mathbf{x}})$.

Nonparametric regression on a quantum computer can be reformulated to exploit quantum properties. First, classical data $\vec{x}$ is encoded into a quantum state
$\ket{\psi_{\vec{x}}}$, which exploits the representation power of feature Hilbert space with highly nonlinear feature map~\cite{mitarai_18,havlíček_19,schuld_19}. The similarity between two data is defined as $K_{m_1,m_2}=\braket{\psi_{\vec{x}^{(m_1)}}}{\psi_{\vec{x}^{(m_2)}}}$. Second, training and prediction can be performed on superposed quantum states of all training data.  
To illustrate this idea, we take a superposition of the training dataset
$\{\mathbf{x}^{(m)}\}\rightarrow M^{-\frac{1}{2}}\sum_m \ket{m}\ket{\psi_{\vec{x}^{(m)}}}\equiv\ket{\psi_{\vec{A}}}$,
$\{y^{(m)}\} \rightarrow  |\vec{y}|^{-\frac{1}{2}}\sum_m y^{(m)}\ket{m}\equiv \ket{\vec{y}}$. The prediction is done by evaluating an overlapping between two states~\cite{schuld_16,zhang_19}: the query state for a set of new data, $\vec{y}\otimes\tilde{\mathbf{x}}\rightarrow \ket{\psi_{R}}\equiv\ket{\vec{y}}\ket{\psi_{\tilde{\mathbf{x}}}}$, and a trained state $\ket{\psi_{\vec{A}^+}}$ that evolves from $\ket{\psi_{\vec{A}}}$, i.e.,
\begin{equation}\label{eq:quantum_prediction}
\tilde{y}=\braket{\psi_R}{\psi_{\vec{A}^+}}.
\end{equation}
A derivation is shown in Supplemental Material (SM)~\cite{supplementary}. Eq.\eqref{eq:quantum_prediction} represents a quantum version of nonparametric learning, serving as a generalization of quantum linear regression in Ref.~\cite{schuld_16,zhang_19} to nonlinear cases. 

Therein, learning is manifested in a proper trained state $\ket{\psi_{\vec{A}^+}}$. A naive choice of $\ket{\psi_{\vec{A}^+}}=\ket{\psi_{\vec{A}}}$ means all training data has equal weighting, neglecting 
correlations between the training data. 
A wisdom from quantum information is to investigate entanglement structure of the bipartite state $\ket{\psi_{\vec{A}}}$. Correlations between data reflect in a Schmidt decomposition of the training state, $\ket{\psi_{\vec{A}}}=\sum_i \lambda_i\ket{u_i}\ket{v_i}$. 
For a least-square loss in Eq.~\eqref{eq:least-square}, 
the trained state $\ket{\psi_{\vec{A}^+}}=c\sum_i g(\lambda_i)\ket{u_i}\ket{v_i}$, where $g(\lambda)=\frac{\lambda}{\lambda^2+\chi}$~\cite{zhang_19}~(see SM~\cite{supplementary}).
The transformation of Schmidt coefficients $\lambda \rightarrow g(\lambda)$ can be considered as entanglement spectrum transformation~\cite{comment1}, and different choices of $g(\lambda)$ may correspond to different types of regression~\cite{comment2}.
 
\begin{figure}
	\includegraphics[width=1.0\columnwidth]{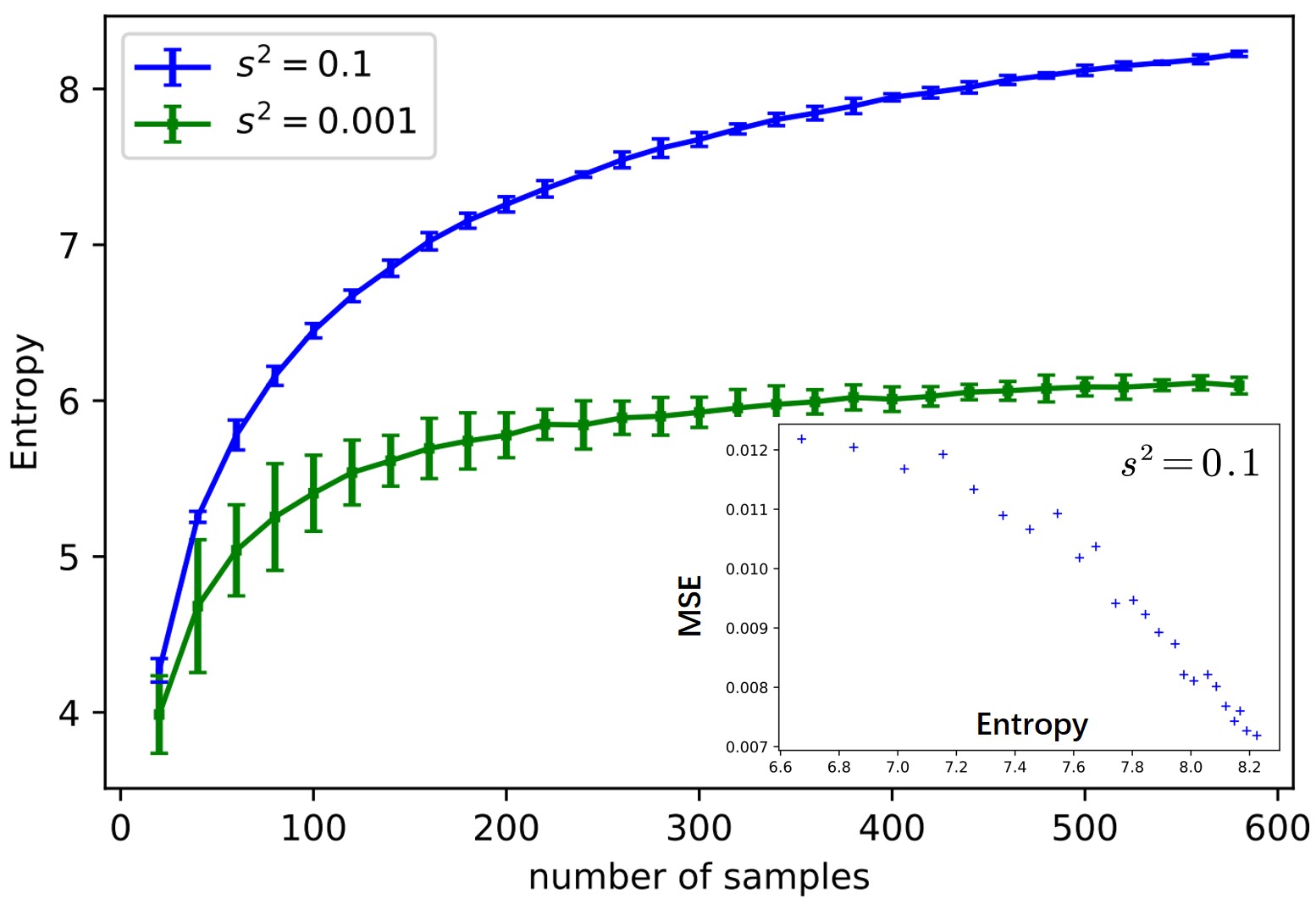}
	\caption{Entanglement entropy 
			{\it vs} the number of samples, where $40$ random datasets are chosen for each set of samples.  The error bars denote standard deviations. 
	The insert shows that the mean-square-error decrease with the entropy.}
	\label{fig:ES_n_samples}
\end{figure}

It is inspiring to investigate the role of entanglement entropy $\mathcal{S}$ of bipartite quantum state $\ket{\psi_{\vec{A}}}$ for machine learning. For illustration, we use squeezing-state encoding with a varied squeezing factor $s$ for the Boston dataset~\cite{scikit-learn}. The similarity function between two samples is $K_{m_1,m_2}=e^{-s^2|\vec{x}^{(m_1)}-\vec{x}^{(m_2)}|^2}$, and samples are less distinguishable for smaller $s$. As seen from Fig.~\eqref{fig:ES_n_samples},  $\mathcal{S}$ increases with the number of samples and saturates faster for smaller $s$. Moreover, the mean-square error decreases with $\mathcal{S}$, indicating that the entanglement entropy may be related to the model capacity that quantifies the ability to fit complicated data~ (see SM~\cite{supplementary}).

\emph{Matrix inversion.--} An efficient quantum algorithm can be developed to obtain $\ket{\psi_{\vec{A}^+}}$ from $\ket{\psi_{\vec{A}}}$. Note that the covariance matrix can be evaluated as $\rho_{K}=K/\text{Tr}K=\text{Tr}_{1}\ket{\psi_\vec{A}}\bra{\psi_\vec{A}}$(partial trace of the addressing registers $\ket{m}$) and $\vec{I}\otimes\rho_{K}\ket{\psi_\vec{A}}= \lambda_i^2\ket{\psi_\vec{A}}$.  The required evolution operator is given by
\begin{equation}\label{evolution_operator}
\ket{\psi_{\vec{A}^+}}=\vec{I}\otimes\vec{B}^{-1}\ket{\psi_{\vec{A}}}.
\end{equation}
where $\vec{B}=\rho_{K}+\chi\vec{I}$.

The non-unitary operator $\vec{B}^{-1}$ is a matrix inversion and its quantum algorithm can exhibit exponential speed-up.  We take an approach for the matrix inversion of $\vec{B}$ by writing it into a combination of unitary operators~\cite{childs_17,arrazola_18}. Inspired by $b^{-1}=\int_{-\infty}^{\infty}dx\delta(bx)=\int_{-\infty}^{\infty}dxdy\exp(ibxy)$, we consider $\vec{B}\ket{b}=b\ket{b}$,  we have
\begin{eqnarray}\label{matrix_inversion}
\vec{B}^{-1}&=&\int_{-\infty}^{\infty}dq_xdq_y\exp(i\vec{B}q_xq_y)\nonumber \\
&\propto& \bra{0_{p_x}}\bra{0_{p_y}}\exp(i\vec{B}\hat{q_x}\hat{q_y})\ket{0_{p_x}}\ket{0_{p_y}},
\end{eqnarray}
where $\ket{0}_p$ is zero momentum eigenstate.
It can be seen that $\vec{B}^{-1}$ can be written as an average of unitary operator $\exp(i\vec{B}\hat{q_x}\hat{q_y})$ over the infinite squeezing state
$\ket{0_{p_x}}\ket{0_{p_y}}$ of momentums $p_x$ and $p_y$.

The state transformation $\ket{\psi_{\vec{A}}}\rightarrow\ket{\psi_{\vec{A}^+}}$ can be implemented as follows:
$\vec{B}^{-1}$ performs on the initial state $\ket{\psi_{\vec{A}}}\ket{0_{p_x}}\ket{0_{p_y}}$, and then project two qumodes onto $\ket{0_{p_x}}\ket{0_{p_y}}$. To implement $\vec{B}^{-1}$, we can write $\exp(i\vec{B}\hat{q_x}\hat{q_y})=\exp(i\rho_K\hat{q_x}\hat{q_y})\exp(i\chi\hat{q_x}\hat{q_y})$. The first part $\exp(i\rho_K\hat{q_x}\hat{q_y})$ can be generated by density matrix exponentiation by sampling from multiple copies of quantum state $\ket{\psi_{\vec{A}}}$~\cite{lloyd_14,kimmel_17}. The second part is just a basic two-qumode gate.

\begin{figure}	
	\includegraphics[width=1.0\columnwidth]{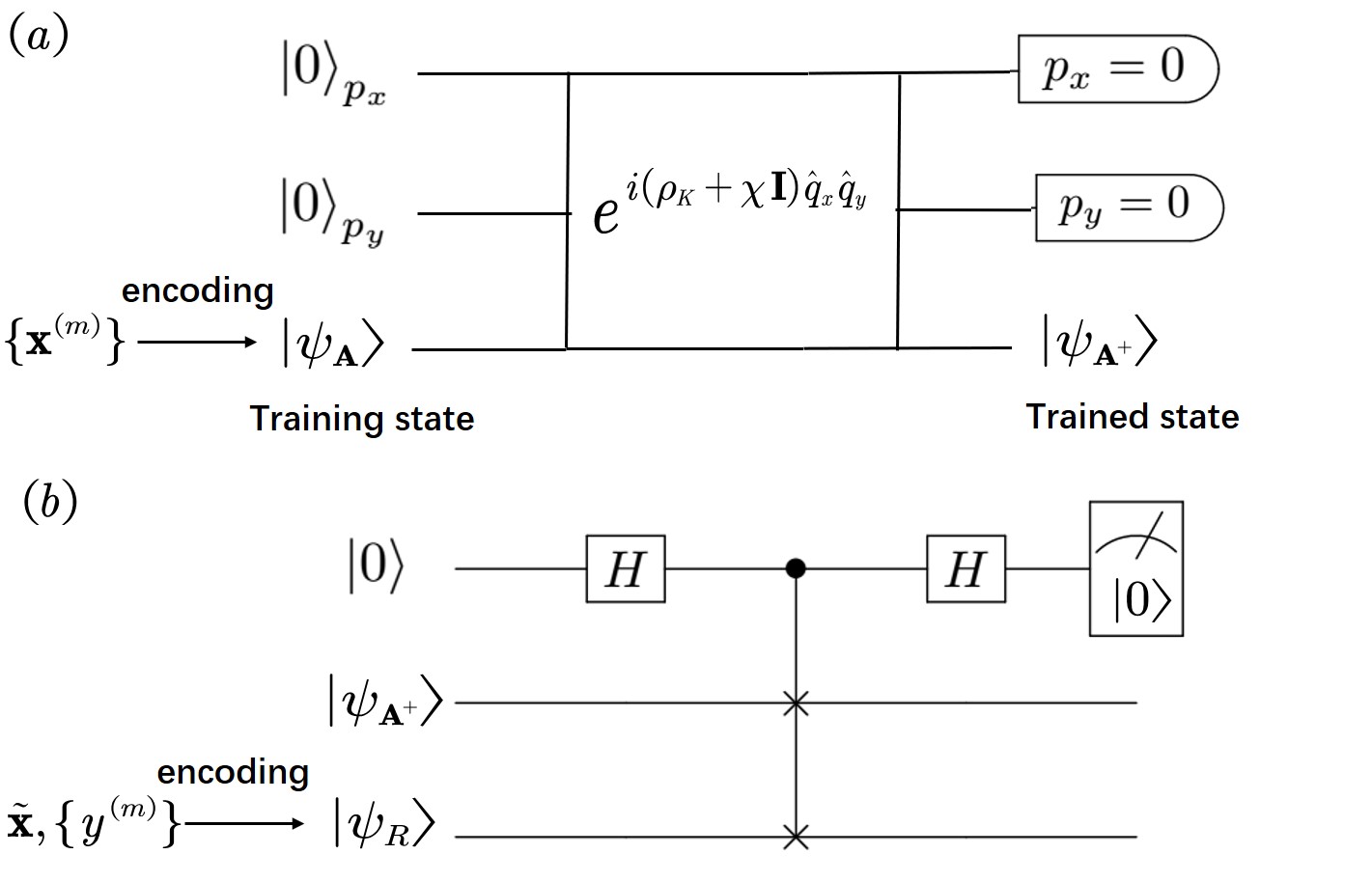}
	\caption{Illustration of the quantum algorithm. (a). Matrix inversion algorithm for a matrix $B=\rho_K+\chi\vec{I}$ that transform $\ket{\psi_\vec{A}}$ into $\ket{\psi_{\vec{A}^+}}=B^{-1}\ket{\psi_{\vec{A}}}$, using two auxiliary qumodes that are post-selected into zero momentum. (b). A swap test that evaluates the inner product between $\ket{\vec{A}^+}$ and $\ket{\psi_R}$, which can be used to infer the prediction for input data $\tilde{\vec{a}}$. } \label{sheme_algorithm}
\end{figure}

\emph{Quantum algorithm.--} We now turn to work out a quantum algorithm for nonparametric regression, basically following techniques in Ref.~\cite{zhang_19}. The main steps are show in Fig.~2, where steps $1-4$ illustrated in Fig.~2a  transform $\ket{\psi_{\vec{A}}}$ to $\ket{\psi_{\vec{A}^+}}$,  and step $5$  illustrated in Fig.~2b implements the prediction.

\emph{1. State preparation}. Prepare the data state $\ket{\psi_{\vec{A}}}$, the query state $\ket{\psi_R}$, and a two-qumode state $\ket{s}_{p_x}\ket{s}_{p_y}$, where $\ket{s}_p=s^{1/2}\pi^{-1/4}\int dp e^{-s^2p^2/{2}}\ket{p}_{p}$.  $\ket{\psi_{\vec{A}}}$ can be prepared efficiently with a quantum random access memory~\cite{giovannetti_08}. It uses the addressing state $\sum_m\ket{m}$ to access the memory cells storing quantum states $\ket{\phi_{\vec{x}^{(m)}}}$ in training data registers. Also, two qumodes are initialed in a finite squeezing state $\ket{s}_{p_x}\ket{s}_{p_y}$.
	
\emph{2. Quantum phase estimation}. Perform  $U=\exp(i\rho_K\hat{q_x}\hat{q_y})$ on $\ket{\psi_\vec{A}}\ket{s}_{p_x}\ket{s}_{p_y}$, where $U$ is constructed with the density matrix exponentiation method~\cite{lloyd_14,kimmel_17,lau_17}. 
The quantum state becomes
\begin{eqnarray}\label{eq:intermediate_state1}
	\sum_i\int dq_xdq_y\frac{\lambda_i e^{-(q_x^2+q_y^2)/2s^2}}{s}\ket{u_i}\ket{v_i}\ket{q_x}_{q_x}\ket{q_y}_{q_y}e^{i\lambda_i^2q_xq_y} \nonumber\\
\end{eqnarray}

\emph{3. Regularization}. Perform $e^{i\chi \hat{q_x}\hat{q_y}}$ on two qumodes. Here $\chi$ is a preset hyperparameter. The state is the same as Eq.\eqref{eq:intermediate_state1} by changing the phase factor to $e^{i(\lambda_i^2+\chi)q_xq_y}$.  

\emph{4. Singular-value transformation}. Project two qumodes into the squeezing state $\ket{s}_{p_x}\ket{s}_{p_y}$, and the state turns to be $\ket{\psi'_{\vec{A}^+}}=\sum_i f(\lambda_i,s,\chi)\ket{u_i}\ket{v_i}$, approximating $\ket{\psi_{\vec{A}^+}}$, where
$f(\lambda_i,s,\chi)=\frac{\lambda_i}{\sqrt{\frac{4}{s^4}+(\lambda_i^2+\chi)^2}}$.

\emph{5. Prediction}. For new data $\tilde{\vec{x}}$, the prediction  $\tilde{y}\propto\braket{\psi_R}{\psi'_{\vec{A}^+}}$ can be accessed with a swap test. After the conditional swap operation, an entangled state is obtained,
$\ket{\Psi}=\frac{1}{\sqrt{2}}(\ket{0}\otimes\ket{\psi'_{\vec{A}^+}}\ket{\Psi_R}+\ket{1}\otimes\ket{\Psi_R}\ket{\psi'_{\vec{A}^+}}$. Then, a Hadamard gate is performed on the qubit, followed with a projection into $\ket{0}$, whose success rate $p=\frac{1}{2}(1+|\braket{\psi'_{\vec{A}^+}}{\psi_R}|^2)$ is used to infer the prediction $\tilde{y}\propto\sqrt{2p-1}$, up to a sign.

\emph{Quantum advantages.--} We now elaborate that the above algorithm has an exponential speed-up. Using quantum random access memory $\ket{\psi_\vec{A}}$ can be prepared in a runtime of $O
(\log M)$. It takes $O(\varepsilon^{-1})$ copies of $\ket{\psi_\vec{A}}$, thus a runtime of $O(\varepsilon^{-1}\log M)$ to perform $\exp(i\rho_K\hat{q_x}\hat{q_y})$~\cite{lau_17,zhang_19}, for a desired accuracy $\varepsilon$. The success rate of homodyne detection is $O(s^{-4})$ and this procedure thus requires $O(s^4)$~(see SM~\cite{supplementary}). In total, the runtime scales as $O(s^4\varepsilon^{-1}\log M)$.
The exponential speed-up relies on the capacity of superposition. If   randomly chosen $M'<M$ training data is superposed for each copy~\cite{rebentrost_18}, then the number of copies should be increased $O(\frac{M}{M'})$ times. To retain exponential speed-up requires $M'/M\sim O(1)$.

Another potential quantum advantage comes from quantum feature map when encoding $\vec{x}$ into $\ket{\phi_\vec{x}}$. Remarkably, continuous variable provides infinite dimension Hilbert space with highly nonlinear feature maps. For instance, encoding into a Gaussian state, such as $\ket{\phi_\vec{x}}=\otimes_i\ket{\vec{x}_i}_c$ ($\ket{\vec{x}_i}_c$ denotes a coherent state with a displacement $\vec{x}_i$), corresponds to a Gaussian kernel, since $\braket{\phi_\vec{u}}{\phi_\vec{v}}=e^{-|\vec{u}-\vec{v}|^2/2}$. Classically intractable instantaneous quantum polynomial or continuous variable instantaneous quantum polynomial circuits are pursued~\cite{bremner_2016,douce_2017}. Moreover, a promising direction is to find encoding schemes that can better represent similarities between data for specified tasks, and thus require less training data and better generalization, such as predicting ground state energies for molecules~\cite{rupp_12,zhang_18}.

\emph{Quantum operations required in trapped ions.--}
Implementing the above quantum algorithm requires  hybrid discrete and continuous variable quantum computing. Some promising candidates for quantum computation, such as superconducting qubits in a circuit-QED and trapped ions, have this property. Here we take trapped ions as the platform~\cite{leibfried_03,häffner_08,monroe_13} to illustrate the details.
 We consider trapped ions in a Paul trap, and take $L$ internal levels of each ions as a qudit to encode the discrete variables and local transverse phonon modes (along $x$ and $y$ directions)~\cite{zhu_06,shen_14} to encode the continuous variables, while the longitudinal collective modes along $z$ direction serves as the bus modes to connect any two ions. Notably
both internal states and phonon modes are well controllable in trapped ions~\cite{cirac_95,lamata_07,gerritsma_10,lau_12,shen_14,ortiz-gutiérrez17,Zhangjunhua_2018,flühmann_19}.

We outline quantum operations required for the proposed algorithm (see SM~\cite{supplementary}). We first address the operations acting on single ion, denoting as the $j$-th ion. A single qubit gate $\mathcal{R}(\theta,\vec{n})=e^{i\theta\sigma_{j\vec{n}}}$ acting on any two internal levels of the $j$-th ion with high fidelity is realizable, where $\sigma_{j\vec{n}}$ is a Pauli matrix along the direction $ \vec{n}$. Operations on a motional mode include $\mathcal{P}_{\alpha}(\theta)=e^{i\theta a_{j\alpha}^\dagger a_{j\alpha}}$, displacement operator $\mathcal{D}_{\alpha}(h)=e^{h a_{j\alpha}-h^*a_{j\alpha}^\dagger}$ and squeezing operator $\mathcal{S}_{\alpha}(s)=e^{-\frac{\ln s}{2}(a_{j\alpha}^2-a_{j\alpha}^{\dagger2})}$ with ${\alpha}=x,y$ ~\cite{cirac_93,leibfried_03,lau_12,kienzler_15,burd_18}, where $a_{j\alpha}  ^{\dagger}$ $(a_{j\alpha})$  is the create (annihilation) operator of the $\alpha$ phonon mode.
A controlled phase gate $\mathcal{C}_q=e^{i\chi\hat{q}^x_j\hat{q}^y_j}$ coupling both motional modes can be realized by manipulating the trap potential. By using red and blue side excitations induced by lasers, internal and motional states can be coupled, e.g., obtaining Dirac type operators $H_1=g\hat{q}^x_j\sigma^x_j$ and $H_2=g\hat{q}^y_j\sigma_{j}^y$~\cite{lamata_07,gerritsma_10}. Then the hybrid operator $\mathcal{W}(\eta)=e^{i\eta\sigma_{jz} \hat{q}^x_j\hat{q}^y_j}$, which is important for quantum phase estimation, can be constructed by repeatedly applying $1/(g^2\delta t^2)$ times of the quantum evolution $e^{iH_2\delta t}e^{iH_1\delta t}e^{-iH_2\delta t}e^{-iH_1\delta t} =e^{-[H_1,H_2]\delta t^2}+O(\delta t^3)$.

As for two ions, besides the standard controlled-NOT gate~\cite{cirac_95}, a beam-splitter defined as $\mathcal{B}(\theta)=e^{i\theta(a_{j\alpha}^\dagger a_{j+1\alpha}+a_{j+1\alpha}^\dagger a_{j\alpha})}$ is needed, and it was theoretically proposed~\cite{shen_14} and then experimentally achieved recently ~\cite{toyoda_15}.  These two operators thus couple qubit states or qumodes from different ions. Furthermore, a coupling of one qubit from an ion and a qumode from another ion is possible with Dirac type Hamiltonians where spin and momentum (position) come from different ions. Necessary quantum operations on three ions includes controlled swap operators, for which one ion provides a qubit to control a swap for other two ions, either on internal states or motional states. The former has been realized experimentally in trapped ions~\cite{linke_18}.
 On the other hand, precision measurement can be implemented for both qubits~\cite{leibfried_03} and qumodes~\cite{poyatos_96}. Those unitary operators and measurements serve as building blocks for the quantum algorithm of nonparametric regression as well as other hybrid quantum information processing tasks.

\emph{Physical implementation with trapped ions.--}
We illustrate the implementation with a simple example. We just take one ion to encode the training dataset, that is, using only one ion to represent one copy of state $\ket{\psi_{\vec{A}}}$. To this end, we choose $L (=M)$ internal levels of the ion as a qudit to encode the M points dataset and two local transverse phonon modes (along $x$ and $y$ directions) to encode the continuous variables (N=2).


The implementation needs four types of ions which we denote as $a,b,c,d-$ ions.
1). An $a$-ion provides a qubit and two qumodes as auxiliary modes.
2). A $b$-ion is used to store the state $\ket{\psi_{\vec{A}}}$ that encodes all data.  $L$ internal levels and  two local motional modes along the $x,y$ directions are used.  On this ion the state will be transformed into the target state $\ket{\psi_{\vec{A}^+}}$.
3). Several $c$-ions, the number of which depends on the accuracy required for the algorithm, are used for constructing the unitary operator $U$ on the b-ion.  Each $c$-ion is initialized in the state $\ket{\psi_{\vec{A}}}$.
4). A $d$-ion encodes input data for prediction into quantum state $\ket{\psi_R}$.

\begin{figure*}[htbp] \centering
	\includegraphics[width=15cm]{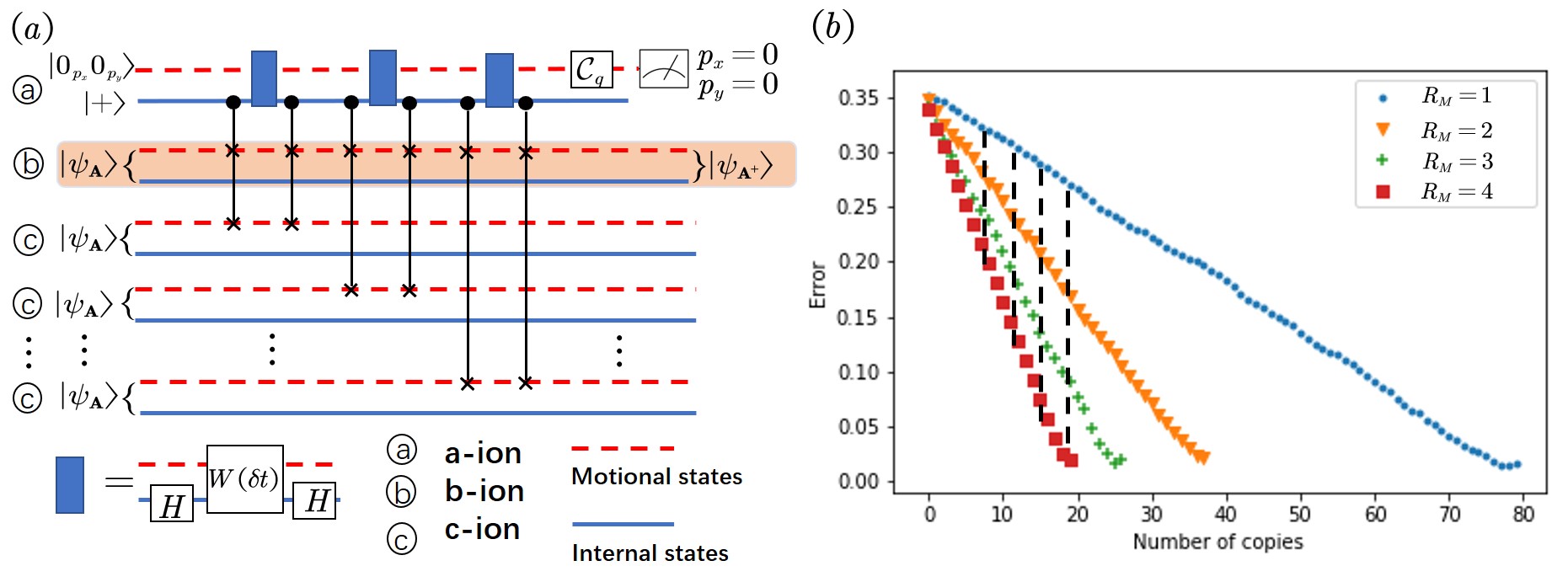}
	\caption{(a). A quantum procedure that transforms $\ket{\psi_\vec{A}}$ into $\ket{\psi_{\vec{A}^+}}$ on $b$-ion, assisted by an $a$-ion providing a qubit for control and two qumodes for matrix inversion, and many $c$-ions initialized in $\ket{\psi_\vec{A}}$ serving as quantum software states for quantum phase estimation. Note the swap only performs on motional states (red dash lines).  (b). Density matrix exponentiation where partial training dataset is superposed. Here $R_M=1,2,3,4$ stands for the number of randomly chosen samples for each copy. The region with dashed lines represents accessible zone for a trade-off between  error and the number of ions involved, constraint by the maximum available c-ion $N_t=20$.} \label{demo_one_ion}
\end{figure*}

The scheme for nonlinear regression is schematically shown in Fig.~3a.
In the state preparation, the generalized Schrodinger cat states  $\ket{\psi_{j\vec{A}}}=\sum_{m=0}^{M}\ket{m_j}\otimes\ket{\phi_j(\vec{x}^{(m)})}$,
where $\ket{\phi_j(\vec{x}^{(m)})}=\ket{\vec{x}^{(m)}_{jx}}\ket{\vec{x}^{(m)}_{jy}}$ for both $j=b,c-$ions, can be generated with Dirac type operations (see SM~\cite{supplementary}). Also two qumodes of the $c$-ion are prepared in a squeezing state $\ket{s}_{p_x}\ket{s}_{p_y}$.
For the quantum phase estimation, the unitary operation $U=\exp(i\rho_K\hat{q_x}\hat{q_y})$ is constructed with the density matrix exponentiation method~\cite{lloyd_14,lau_17,kimmel_17,zhang_19},
\begin{eqnarray}\label{sample-Ham}
&\text{Tr}_{\rho}(e^{i\delta t \hat{q}_x\hat{q}_y S_{cv}}\rho\otimes\rho' e^{-i\delta t\hat{q}_x\hat{q}_y S_{cv}})\nonumber\\
&= e^{i\rho\delta t\hat{q}_x\hat{q}_y}\rho'e^{i\rho\delta t\hat{q}_x\hat{q}_y} +O(\delta t^2).
\end{eqnarray}
Here $\rho=\rho_K$ is a mixed state encoded in the motional states of $c$-ion (the internal states are traced out), and $\rho'\equiv\ket{\psi_\vec{A}}\bra{\psi_\vec{A}}$ is a state
on $b$-ion. The conditional swap operator $e^{i\delta t\hat{q}_x\hat{q}_yS_{cv}}$ is constructed from $\mathcal{C}_{S_{cv}}H_a\mathcal{W}_a(\delta t)H_a\mathcal{C}_{S_{cv}}$~\cite{zhang_18}, where $\mathcal{C}_{S_{cv}}$ swaps motional states of $b$-ion and $c$-ion, conditioned on the qubit state of $a$-ion initialized in $\ket{+}$ state. The one-qubit-two-qumodes coupling $\mathcal{W}_a(\delta t)$ performs on the $a$-ion, and $H_a$ is a Hadamard gate acting on the $a$-ion.
Multiple copies of $c$-ion are required and each is encoded with mixed state $\rho$ in the internal states. Conditional swap operations are sequentially performed on $b$-ion and a new $c$-ion and swap their motional states, effectively giving a $U$ operation on $b$-ion.

After applying $U$ on the $b$-ion, a regularization can be realized by applying $\mathcal{C}_q=e^{i\chi\hat{q}_x\hat{q}_y}$on the two motional modes of $c$-ion.
A measurement  projects two qumodes of the $c$-ion onto $\ket{s}_{p_x}\ket{s}_{p_y}$. The $b$-ion is on target state $\ket{\psi_{\vec{A}^+}}$. After an evolution $U_R^\dagger$, where $U_R\ket{0}=\ket{\psi_R}$, a projective measure on $\ket{g}\ket{0}_c$ with the success probability $p=|\braket{\psi_R}{\psi_{\vec{A}^+}}|^2$ can infer the prediction for new data $\tilde{x}$.

This implementation scheme can demonstrate a remarkable quantum-enhanced property. The above density matrix exponentiation can use partial training dataset for each time~\cite{rebentrost_18}, e.g., use $\rho\sim\sum_{m\in \mathcal{R}_{M}}\ket{\phi_j(\vec{x}^{(m)})}\bra{\phi_j(\vec{x}^{(m)})}$, where $\mathcal{R}_{M}$  represents to randomly choose $R_M$ samples in the training dataset, and we thus choose $L=R_{M}$ internal levels to represent state $\ket{\psi_{\vec{A}}}$.
Therefore, in the experiments, we can compare the results of $R_{M}=1,2,\cdots,M$ randomly chosen data from the dataset for each copy. We calculate the prediction errors as a function of the number of the $c$-ions, and the results are shown in Fig.~3b. 
Under the condition of same accuracy, the number of $c$-ions increases with the decrease of $R_{M}$; similarly, the prediction errors decrease for a large $R_{M}$. 
Therefore, it is a clear evidence to demonstrate the power of superposition for quantum nonparametric learning. A remarkable result presented here is that, a Paul trap with around ten ions, which has been realized in several groups~\cite{zhang_17,friis_18,kokail_19,wright_19}, can demonstrate the quantum-enhanced property for quantum machine learning. (As for the scheme scalability, it is discussed in  SM~\cite{supplementary}.)

To conclude,
we have illustrated a quantum paradigm of nonparametric learning 
that can fully exploit quantum advantages with realistic physical implementation.
The above-proposed experimental scheme  has paved the way for quantum machine learning. 
\begin{acknowledgments}
	This work was supported by the National Key Research and Development Program of China (Grant
	No. 2016YFA0301800), the National National Science Foundation of China (Grants No. 91636218, No.11474153,and No. U1801661), the Key R\&D Program of Guangdong province (Grant No. 2019B030330001), and the Key Project of Science and  Technology of Guangzhou (Grant No. 201804020055).
\end{acknowledgments}

%
\end{document}